\documentclass[%
 reprint,
 amsmath,amssymb,
 aps,
]{revtex4-2}

\usepackage[pdftex]{graphicx}
\usepackage{color}
\usepackage{dcolumn}
\usepackage{bm}
\usepackage{lineno}
\usepackage{comment}
\usepackage{amsmath}
\usepackage{mathtools}
\usepackage{physics}

\newcommand{\mDP}{m_\text{DP}}
\newcommand{\PDP}{P_\text{DP}}
\newcommand{\Pout}{P_\text{out}}
\newcommand{\Pin}{P_\text{in}}
\newcommand{\Trx}{T_\text{rx}}
\newcommand{\Tin}{T_\text{in}}
\newcommand{\Tsys}{T_\text{sys}}
\newcommand{\Aeff}{A_\text{eff}}
\newcommand{\Dnu}{\Delta\nu}
\newcommand{\DnuDP}{\Delta\nu_\text{DP}}
\newcommand{\GA}{B}
\newcommand{\plocalmin}{p^\text{min}}
\newcommand{\pglobal}{p_\text{global}}
\newcommand{\vE}{v_\text{E}}
\newcommand{\kB}{k_\text{B}}

\newcommand{\ueV}{\mu\text{eV}}
\newcommand{\GeV}{\text{GeV}}

\newcommand{\kHz}{\text{kHz}}

\newcommand{\GHz}{\text{GHz}}
\newcommand{\mm}{\text{mm}}

\newcommand{\sqcm}{\text{cm}^2}
\newcommand{\cubcm}{\text{cm}^3}
\newcommand{\K}{\text{K}}
\newcommand{\kmpers}{\text{km}/\text{s}}
\newcommand{\Deg}{^{\circ}}
\newcommand{\aW}{\text{aW}}

\newcommand{\eq}[1]{Eq.~(\ref{eq:#1})}

\newcommand{\fig}[1]{Fig.~\ref{fig:#1}}
\newcommand{\Fig}[1]{Figure~\ref{fig:#1}}
\newcommand{\tab}[1]{Table~\ref{tab:#1}}
\newcommand{\sect}[1]{Sec.~\ref{sec:#1}}

\newcommand{\App}[1]{Appendix~\ref{app:#1}}

\newcommand{\etal}{\bibfnamefont{\emph{et al.}}}
\newcommand{\JCAP}{J. Cosmol. Astropart. Phys.}
\newcommand{\RSI}{Rev. Sci. Instrum.}

\begin{document}

\title{Search for Dark Photon Dark Matter in the Mass Range
$\mathbf{41\text{--}74}\,\boldsymbol{\mu}\mathbf{eV}$ 
\\ using Millimeter-Wave Receiver and Radioshielding Box}

\author{S.~Adachi}
\email{adachi.shunsuke.5d@kyoto-u.ac.jp}
\affiliation{Hakubi Center for Advanced Research, Kyoto University, Kyoto 606-8501, Japan}
\altaffiliation[Also at ]{Department of Physics, Faculty of Science, Kyoto University, Kitashirakawa Oiwake-cho, Sakyo-ku, Kyoto 606-8502, Japan}

\author{R.~Fujinaka}
\affiliation{Department of Physics, Faculty of Science, Kyoto University, Kyoto 606-8502, Japan}

\author{S.~Honda}
\affiliation{Division of Physics, Faculty of Pure and Applied Sciences, University of Tsukuba, Ibaraki, 305-8571, Japan}
\altaffiliation[Also at ]{Tomonaga Center for the History of the Universe (TCHoU), University of Tsukuba, Japan}

\author{Y.~Muto}
\affiliation{Department of Physics, Faculty of Science, Kyoto University, Kyoto 606-8502, Japan}

\author{H.~Nakata}
\affiliation{Department of Physics, Faculty of Science, Kyoto University, Kyoto 606-8502, Japan}

\author{Y.~Sueno}
\affiliation{Department of Physics, Faculty of Science, Kyoto University, Kyoto 606-8502, Japan}

\author{T.~Sumida}
\affiliation{Department of Physics, Faculty of Science, Kyoto University, Kyoto 606-8502, Japan}

\author{J.~Suzuki}
\affiliation{Department of Physics, Faculty of Science, Kyoto University, Kyoto 606-8502, Japan}

\author{O.~Tajima}
\affiliation{Department of Physics, Faculty of Science, Kyoto University, Kyoto 606-8502, Japan}

\author{H.~Takeuchi}
\affiliation{Department of Physics, Faculty of Science, Kyoto University, Kyoto 606-8502, Japan}

\collaboration{DOSUE-RR Collaboration}

\date{\today}

\begin{abstract}
Dark photons have been considered potential candidates for dark matter. 
The dark photon dark matter (DPDM) has a mass and interacts with electromagnetic fields via kinetic mixing with a coupling constant of $\chi$.
Thus, DPDMs are converted into ordinary photons at metal surfaces. 
Using a millimeter-wave receiver set in a radioshielding box, we performed experiments to detect the conversion photons from the DPDM
in the frequency range $\text{10--18}\,\GHz$, which corresponds to a mass range $\text{41--74}\,\ueV$.
We found no conversion photon signal in this range and set the upper limits to $\chi < (0.5\text{--}3.9) \times 10^{-10}$ at a 95\% confidence level.
\end{abstract}

\maketitle

\section{Introduction}

The study on dark matter is a prominent topic in the fields of particle physics and cosmology.
Dark matter is localized in most galactic halos. 
However, its interactions with ordinary particles, standard-model particles, remain unexplained except via gravity.
Dark photons are considered potential candidates for dark matter. 
The mass ($m_{\rm DP}$) is nonzero and the dark photon interacts with the electromagnetic fields via kinetic mixing with a coupling constant of $\chi$~\cite{Wispy}.
Dark photon dark matter (DPDM) in the $\ueV$ or meV mass ranges have been predicted to exist in the context of high-scale inflation models
and a part of string theories~\cite{PhysRevD.93.103520, Wispy}.

DPDMs convert to standard-model photons at the boundary of the mediums via the kinetic mixing~\cite{AxionLimit}.
Metal surfaces are used for such conversions~\cite{DishAntenna, Tomita, DOSUE-K, QUALIPHIDE, SHUKET}.
Because the speed of DPDM ($v_{\rm DP}$) is considerably low compared to the speed of light ($v_{\rm DP}/c \approx 10^{-3}$), 
the direction of the conversion photons is almost perpendicular to the surface of the plate, within $0.1^{\circ}$~\cite{DishDirection}. 
The frequency of the conversion photons approximately $\nu_0$ roughly corresponds to the mass of the DPDM because of energy conservation, that is, $h\nu_0 \simeq m_{\rm DP} c^2$, where $h$ is the Planck constant.
Owing to the low speed of the DPDM, 
the frequency width of the conversion photons ($\Delta \nu_\text{DP}$) is considerably narrow compared with the peak frequency, 
$\Delta \nu_\text{DP}/\nu_0 \approx 10^{-6}$~\cite{PhysRevLett.51.1415}.
Thus, the conversion photons should be observed as a peak in the frequency spectrum.

The signal power of the conversion photons $\PDP$ is denoted by~\cite{DishAntenna}, 
\begin{eqnarray} \label{eq:chi}
    \PDP &=& (6.4 \times 10^{-2}\,\mathrm{aW}) \times
    \left(\frac{\chi}{10^{-10}}\right)^{2}
    \left(\frac{\Aeff}{10\,\sqcm}\right) \nonumber \\
    && \quad \times\left(\frac{\rho}{0.39\,\GeV/\cubcm}\right)
    \left(\frac{\alpha}{\sqrt{2/3}}\right)^2,
\end{eqnarray}
where $\Aeff$ is the effective antenna aperture, 
$\rho = 0.39\pm0.03\,\text{GeV/cm}^{3}$ is the energy density of the dark matter in the galactic halo~\cite{DMdensity},
and $\alpha$ is a factor determined by the angular distribution of the DPDM field relative to the sensitive polarization axis.
We selected $\alpha=\sqrt{1/3}$ for the case of random distribution using a single-polarization detector~\cite{DishAntenna}. 

The DOSUE-RR~(Dark-photon dark-matter observing system for Un-Explored Radio-Range)
aims to detect the conversion photons of DPDM using millimeter-wave receivers.
We achieved stringent constraints, $\chi<(\text{0.3--2.0})\times 10^{-10}$ in the mass range $74\text{--}110\,\ueV$ by using a cryogenic millimeter-wave receiver~\cite{DOSUE-K}.
Searches using millimeter-wave receivers have the advantage of broad coverage of the mass range.
In this study, 
we switched the explored mass range $\text{41--74}\,\ueV$, corresponding to the frequency range $\text{10--18}\,\GHz$.

The remainder of this paper is organized as follows.
In Sec.~II, we describe the experimental setup.
The calibration procedures are described in Sec.~III. 
Section~IV details the measurement procedure and analysis of the DPDM search. 
Section~V describes the systematic errors.
In Sec.~VI, we present the constraints on the coupling constant $\chi$.
Finally, Sec.~VII presents the conclusions of this study. 

\section{Experimental Setup}

\begin{figure}[htb]
\includegraphics[width=8.6cm]{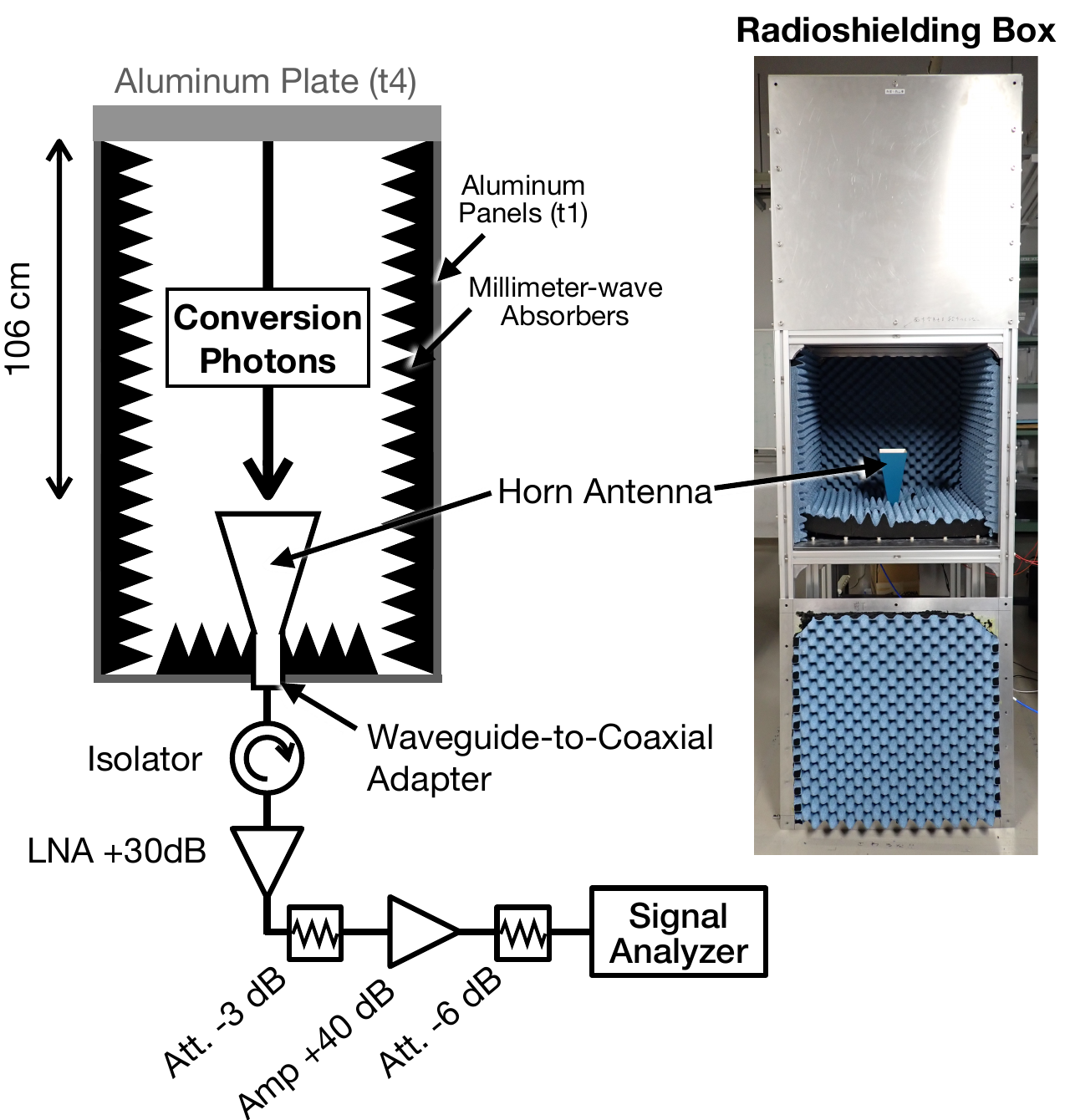}
\caption{\label{fig:setup}
Our setup for the DPDM search.
Optical components were assembled in a radioshielding box.
We closed the front panel of the box when we collected data.
The box comprises aluminum panels and radio-wave absorbers.
We set an aluminum plate and a horn antenna at the top and bottom of the box, respectively.
The conversion photons were collected by the antenna.
The amplified signals were measured by a signal analyzer.
}
\end{figure}

\Fig{setup} shows the setup to search for DPDM.
A radioshielding box was essential for this measurement because manmade millimeter-wave signals and 
their harmonics can enter the antenna, for example, Wi-Fi and telecommunication signals.
This box comprises 1 mm-thick aluminum plates and sheets of the radio-wave absorber, Eccosorb CV-3, with a reflectance of $-40\,\text{dB}$ above 8 GHz~\cite{eccosorb}. 
An aluminum plate with a thickness of 4 mm and an area of $690\,\mm \times 690\,\mm$ was placed at the top of the box.
The conversion photons from the plate surface entered the horn antenna at the bottom of the box.
The horn antenna was Pasternack PE9856/SF-60, with an aperture size of $123.8\,\mm \times 91.9\,\mm$.
Using a tiltmeter, we confirmed that the antenna direction was perpendicular to the plate with $0.1\Deg$ precision.

The signals collected by the antenna were transmitted to a coaxial cable through a waveguide-to-coaxial adapter.
A low-noise amplifier (Low-Noise Factory LNF-LNR6\_20A) and another amplifier (Mini-Circuits ZVA-183G-S+) were used to amplify the signals.
An isolator (Pasternack PE8304) and two attenuators were inserted to suppress standing waves over the signal path.
The amplified signals were measured using a signal analyzer (ANRITSU MS2840A).
The signal analyzer performed a Fast Fourier Transform to measure the frequency spectra. 
The resolution bandwidth was set to 300 Hz ($\ll \DnuDP \approx 10\,\kHz$).
The signal analyzer can simultaneously obtain spectral data for a limited frequency range of 2.5 MHz using this resolution bandwidth.
There are 32,769 data points in each dataset for the 2.5 MHz range, that is, the frequency interval was 76.3 Hz.

\section{Calibration}

\subsection{Effective antenna aperture and beam pattern}\label{sec:Aeff}

\begin{figure}[htb]
\includegraphics[width=8.6cm]{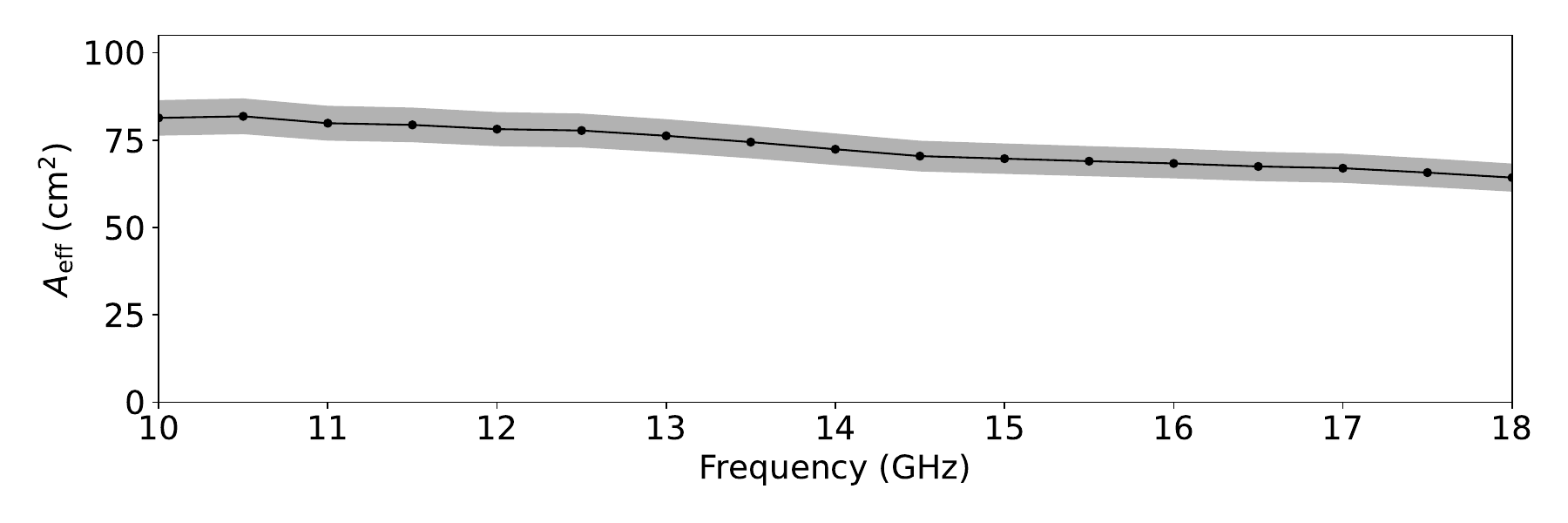}
\caption{\label{fig:Aeff}
Simulated $A_\mathrm{eff}$ as a function of the frequency. 
A gray band indicates the size of the systematic error.
}
\end{figure}

The effective antenna aperture $\Aeff$ 
is a visible area of the antenna and is smaller than its physical aperture area.
This can be estimated from the beam pattern $\GA$ using the antenna theorem~\cite{RadioAstro}:
\begin{eqnarray*}
    \Aeff\Omega_A &=& \lambda^2,  \label{eq:AeffOmega} \\
    \Omega_A &=& \int \GA (\theta, \phi) d\Omega, \label{eq:Omega}
\end{eqnarray*}
where $\lambda$ is the wavelength, $\theta$ is the polar angle, and $\phi$ is the azimuthal angle.
$\GA$ is defined as the relative antenna gain in each direction, whose peak value is normalized to 1.
Using a finite element method simulation software, ANSYS-HFSS (2022R2), 
we simulated the beam patterns from 10.0 GHz to 18.0 GHz with 0.5 GHz intervals.
Thereafter, we calculated the $\Aeff$ at each frequency, as shown in \Fig{Aeff}. 

\begin{figure}[tb]
\includegraphics[width=8.6cm]{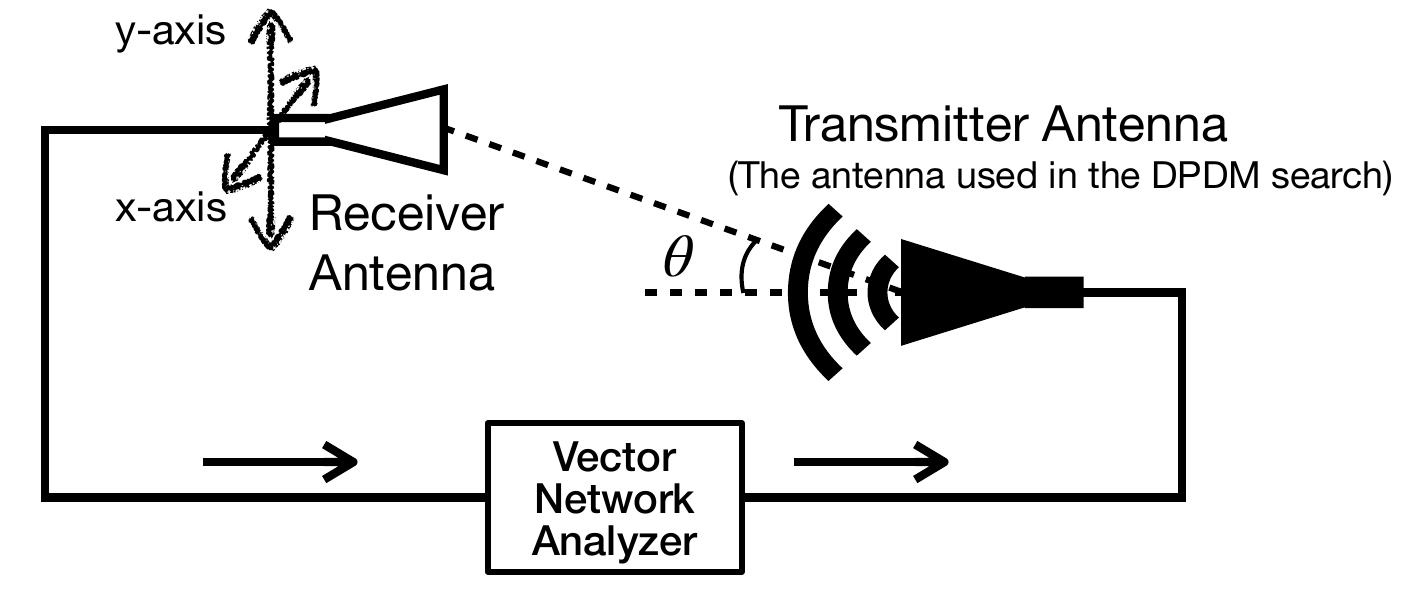}
\caption{\label{fig:beam_setup}
    Conceptual illustration for the setup of the beam pattern measurement.
    A vector network analyzer outputs a monochromatic millimeter-wave signal from a transmitter antenna, 
    which is the horn antenna used in the DPDM search.
    The output signal was collected by a receiver antenna set on the X-Y movable stage. 
    The beam pattern was measured by moving the X-Y stage.
    The beam patterns of the receiver antennas were known in advance 
    and their effects were considered in calculating the beam pattern of the transmitter antenna.
}
\end{figure}

To validate the simulation results,
we measured the beam patterns at each frequency using a near-field measurement system, as shown in \fig{beam_setup}.
The vector network analyzer outputs a monochromatic millimeter-wave signal from the transmitter antenna, 
which is the horn antenna used in the DPDM search.
The output signal was collected using another antenna (the receiver antenna) set on an X-Y movable stage.
Further, the power was measured using the vector network analyzer.
The measurements were repeated by moving the receiver antenna.
The maximum angular coverage of these measurements was $20\Deg$.
The measured frequencies were the same as those in the simulation, that is, from 10.0 GHz to 18.0 GHz with 0.5 GHz intervals. 
We used two types of receiver antenna with the waveguides WR-90 and WR-62 for 10.0--12.0 GHz and 12.5--18.0 GHz measurements, respectively.
The distances between the two antennas were $192\,\mm$ and $238\,\mm$ for measurements with WR-90 and WR-64, respectively. 

\begin{figure}[b]
\includegraphics[width=8.6cm]{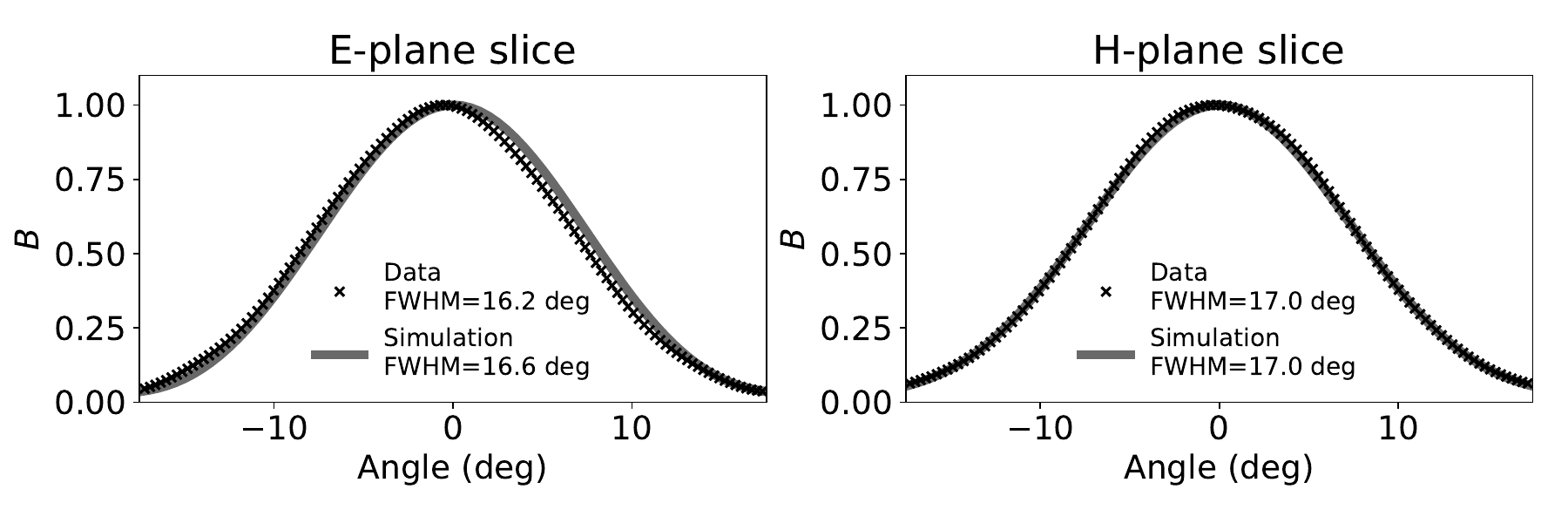}
\caption{\label{fig:beam}
    Beam pattern $\GA$ of the horn antenna at 10 GHz in E-plane (left) and H-plane (right).
    Points and solid lines represent the measurement and simulation, respectively.
}
\end{figure}

\begin{figure}[t]
\includegraphics[width=8.6cm]{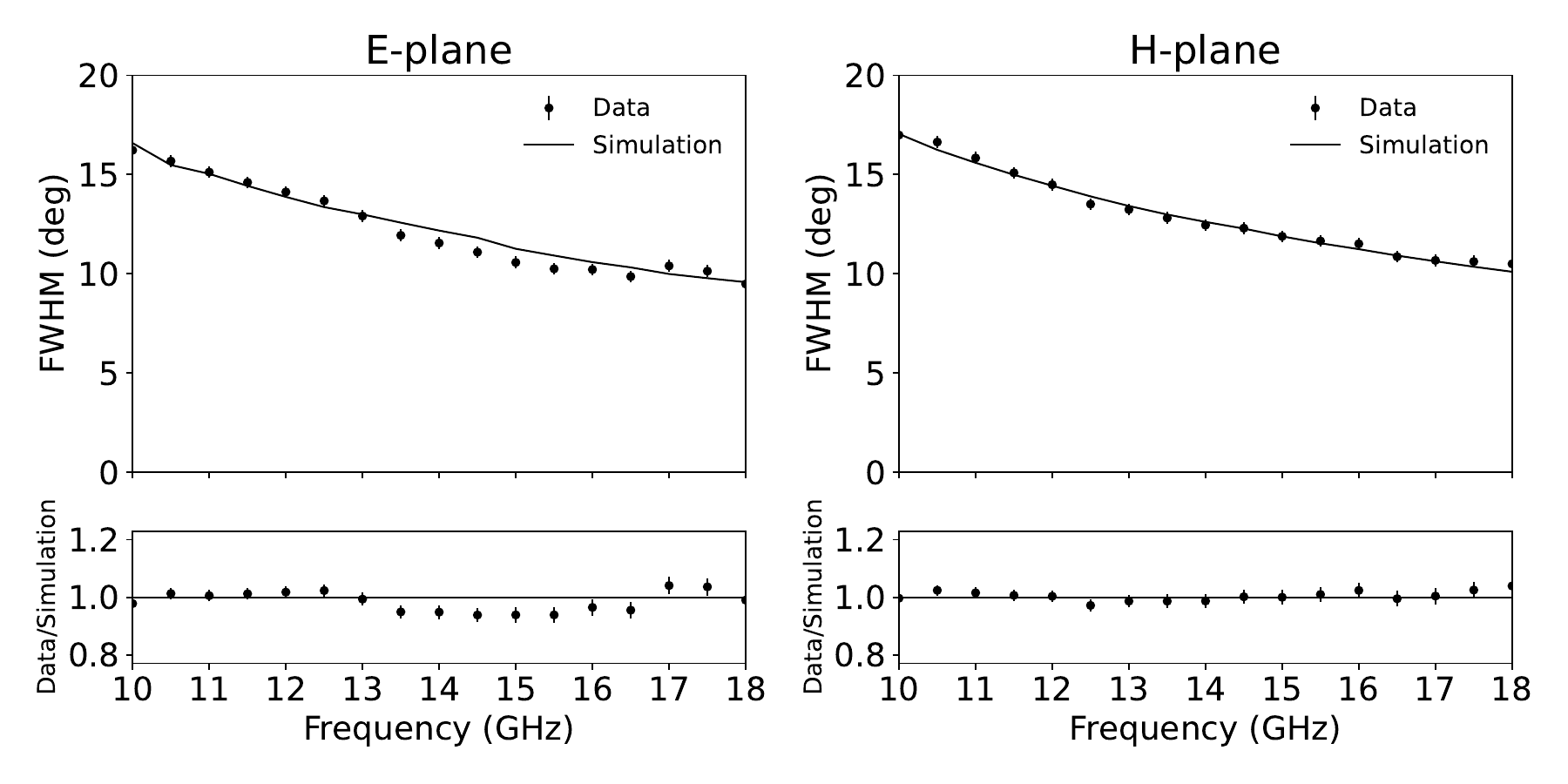}
\caption{\label{fig:beamFWHM}
Beam width, full-width-half-maximum (FWHM) of the beam pattern, 
as a function of frequency in E-plane (left) and H-plane (right).
Points and solid lines indicate the measurement and simulation, respectively.
The bottom panels show their ratios.
The scan step of the X-Y stage dominates the error in the measurement.
}
\end{figure}

The beam pattern measured at 10.0 GHz is shown in \fig{beam}.
The simulated beam pattern was superimposed for comparison.
The beam widths at each frequency are shown in \fig{beamFWHM}.
The difference between the measurement and simulation was less than 6.6\%, 
which is sufficiently small for this study.

\begin{figure}[htb]
\includegraphics[width=8.6cm]{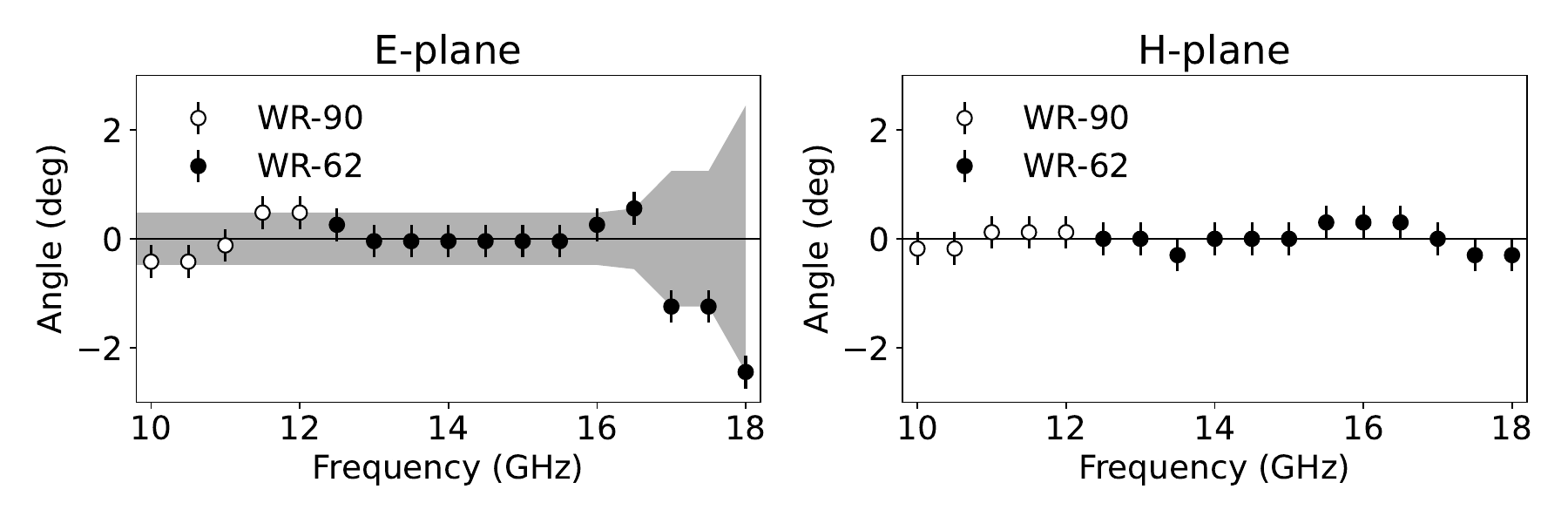}
\caption{\label{fig:beamcenter}
Measured beam centers at each frequency in E-plane (left) and H-plane (right).
The open markers and filled markers indicate the type of receiver antenna used in the measurements.
Shifts from $0\Deg$ were assigned to the systematic error.
A gray band indicates the size of the systematic error.
}
\end{figure}

\Fig{beamcenter} shows the measured beam center at each frequency.
We observed shifts in the beam center from $0\,\Deg$, which did not appear in the simulation.
This was evident above 17 GHz in the E-plane.
The mechanical asymmetry in the E-plane of the waveguide-to-coaxial adapter causes the shifts.
Therefore, a systematic error was assigned.
The gray band in the E-plane of \fig{beamcenter} indicates the size of the systematic error used in the analysis.

As regards $\Aeff$, we evaluated the difference between the simulation and measurement
by using calculations with limited angular coverage within $\text{17.5}\,\Deg$.
The maximum difference between them is $6.2\%$, 
which is assigned as a systematic error, as shown in \fig{Aeff}.

\subsection{Receiver gain}

\begin{figure}[htb]
\includegraphics[width=8.6cm]{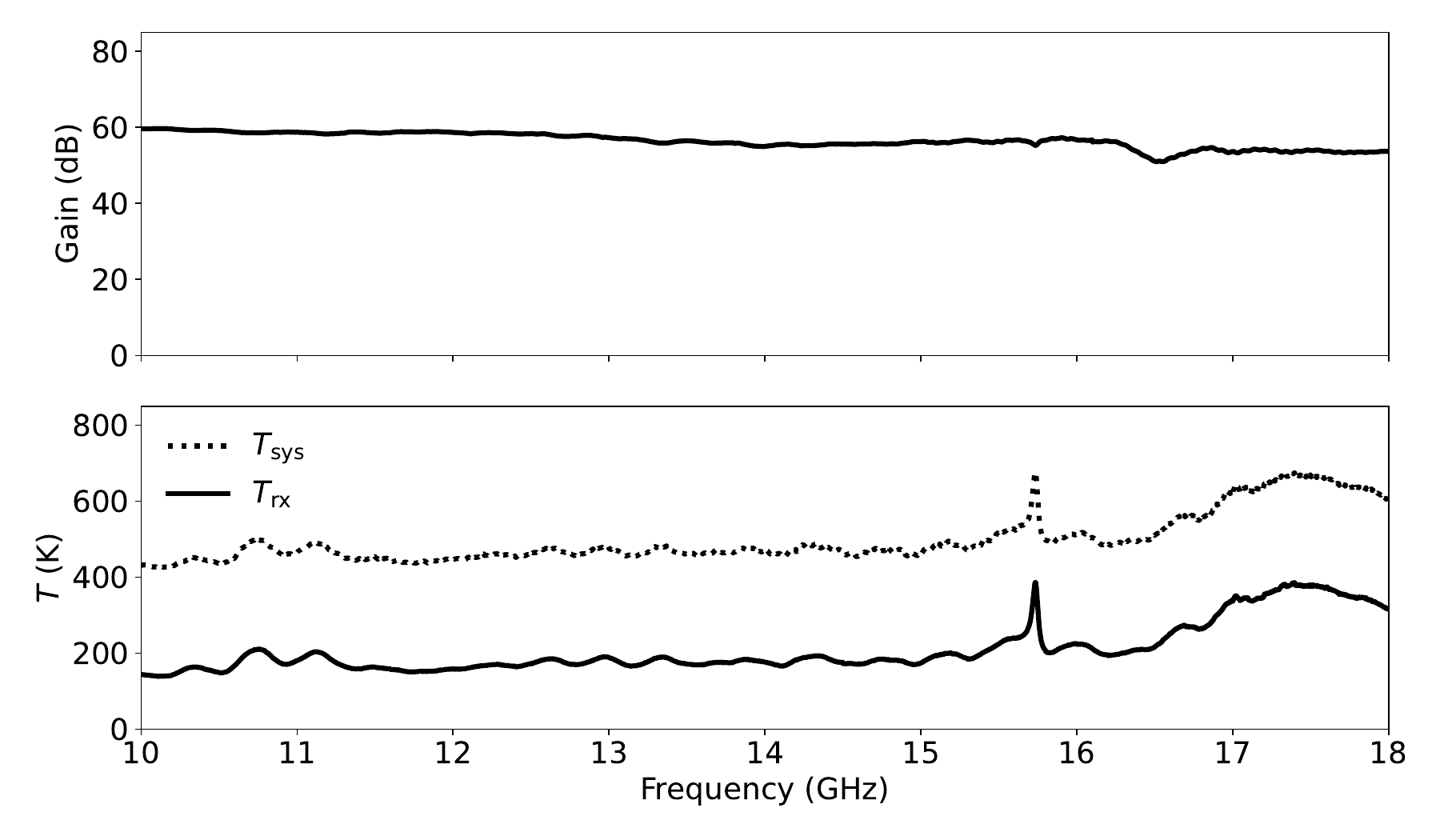}
\caption{\label{fig:gain}
Receiver gain ($G$), receiver temperature ($\Trx$), and system temperature ($\Tsys$) as a function of frequency.
$\Tin \equiv \Tsys - \Trx$ corresponds to loading into the antenna,
which was dominated by the room temperature radiation ($\sim 290\,\K$).
A loss of the waveguide-to-coaxial adapter increases the $\Trx$ at 15.7 GHz.
}
\end{figure}

The power measured by the signal analyzer $\Pout$ is 
an amplified total power of the receiver input power $\Pin$ and receiver noise, 
which is represented by the receiver temperature $\Trx$:
\begin{equation}\label{eq:Pout}
\Pout = G \left( \Pin + \kB \Trx \Dnu \right),
\end{equation}
where $G$ is the receiver gain, $\kB$ is the Boltzmann constant, and $\Dnu$ is the frequency bandwidth.
To obtain $G$ and $\Trx$, we performed a calibration using hot (room temperature) and cold (77K) blackbody radiation sources.
The millimeter-wave absorber Eccosrb CV-3 was used as the blackbody radiation source.
The temperature under room conditions ($T^\text{hot}\sim290\,\K$) was measured using a radiation thermometer for each calibration.
Cold radiation ($T^\text{cold}=77\,\K$) was emitted by an absorber immersed in a liquid-nitrogen bath made of a foamed polystyrene box, 
transparent to radio waves~\cite{quasioptical, RTMLI}.
Each source was set in front of the antenna aperture to fully cover the antenna.
Each $\Pin$ was calculated as $\Pin^\text{hot}=\kB T^\text{hot}\Dnu$ or $\Pin^\text{cold}=\kB T^\text{cold}\Dnu$.
On this basis, we obtained $G$ and $\Trx$ according to \eq{Pout}.
In this calibration, we set the resolution bandwidth to be the same as that used in the DPDM search.
\Fig{gain} shows the $G$ and $\Trx$ as a function of frequency.
The typical gain was $60\,\text{dB}$ and $\Trx$ was $\approx 200\,\K$.
The system temperature is the sum of $\Trx$ and input radiation ($\Tin$) 
during the DPDM search; that is, $\Tsys = \Tin + \Trx$.
$\Tin$ is dominated by the thermal loading at room temperature.

\section{Measurement and analysis}\label{sec:measurement}
The DPDM search was performed from March 6 to 17, 2023. 
We collected data for frequencies ranging from 10.0 GHz to 18.0 GHz, corresponding to a DPDM mass range of 41 $\ueV$ to 74 $\ueV$. 
The time to accumulate data was set to 2 s for each data chunk.
The center frequency was shifted by 2.0 MHz after obtaining 12 chunks of data for each frequency region.
We collected 48,000 data chunks in 4,000 frequency regions.
Two consecutive frequency ranges were overlapped and used to estimate the statistical error in the analysis.
Further, we performed gain calibration before and after each 100 MHz data acquisition.
The typical time interval between the calibrations was 40 min.
The maximum difference in $G$ was $1.5\%$.

The analysis method is the same as that used in our previous study~\cite{DOSUE-K}.
The 12 data chunks at each frequency were averaged and rebinned to $\Dnu = 2\,\kHz$. 
The averaged spectra were converted to $\Pin$ by \eq{Pout}.
To extract the signal power of the conversion photons $\PDP$, 
we fitted a function to the $\Pin$ spectra at each frequency for the signal peak $\nu_0\equiv \mDP c^2/h$.
The fitting function at $\nu_0$ comprises a signal model, $f_\text{sig}$
and the baseline background noise $f_\text{bg}$, which is a one-dimensional polynomial $f_\text{bg} (\nu; a, b) = a (\nu - \nu_0) + b$, 
\begin{equation}\label{eq:fitfunc}
 f(\nu; \PDP, a, b) = \PDP \times f_\text{sig} (\nu; \nu_0) + f_\text{bg}(\nu; \nu_0, a, b)~.
\end{equation}
Here, $a$ and $b$ are the parameters used to model the baseline noise, and
$f_\text{sig}$ is the difference in the cumulative functions introduced to account for the effect of the finite bin width
\begin{equation}\label{eq:fitfsig}
 f_\text{sig} (\nu; \nu_0) = F(\nu + \Dnu/2; \nu_0) - F(\nu - \Dnu/2; \nu_0),
\end{equation}
where,
\begin{eqnarray}\label{eq:fitF}
F(v) &=& \int_{0}^{v}{\text{d}v'}\int^{4\pi}{\text{d}\Omega ~g(\vb{v'};v_\text{c}, \vb{v_{\text{E}}})~v'^2 }~,
\\
    g(\vb{v};v_\text{c}, \vb{v_{\text{E}}}) &=&
    \frac{1}{\left(\sqrt{\pi}v_\text{c}\right)^{3}}\exp{-\frac{|\vb{v}+\vb{v_{\text{E}}}|^2}{v_\text{c}^2}}. \label{eq:fitg}
\end{eqnarray}
Here, $F$ is a function of the DPDM speed; 
$v \equiv \abs{\vb{v}} = c \sqrt{1 - (\nu_0/\nu)^2}$; 
$g(\vb{v})$ denotes the velocity distribution,; 
$v_\text{c}$ denotes the circular rotational speed of the galaxy,
and $\vb{v_{\text{E}}}$ is the velocity of the Earth within the frame of the Galaxy.
We assume a Maxwell-Boltzmann distribution for $g(\vb{v})$~\cite{DMVelocityDistribution}.
Furthermore, we assumed $\abs{\vb{v}_\text{E}} = v_\text{c} = 220\,\kmpers$ in the fittings, as in many dark matter searches~\cite{DMVelocity,LUX,Xenon1T, PandaX,DarkSide}.
The speed of the Earth $\vE$ varied from $220\,\kmpers$ to $242\,\kmpers$ during the search measurement 
owing to the revolution and rotation of the Earth~\cite{DMVelocity}.
This caused a small change in the peak frequency.
This effect is considered as a systematic error in \sect{syst}.

We varied the peak frequency $\nu_0$ from 10.0 to 18.0 GHz at small intervals of $\Delta \nu= 2\,\kHz$
and performed a fit by floating $\PDP$, $a$, and $b$ in the frequency range from $\nu_0 - 50\,\kHz$ to $\nu_0 + 200\,\kHz$.
The error on each data point in the fit was estimated from the standard deviations of the data for the 0.25 MHz regions below and above the fit range. 

\begin{figure}[htb]
\includegraphics[width=8.6cm]{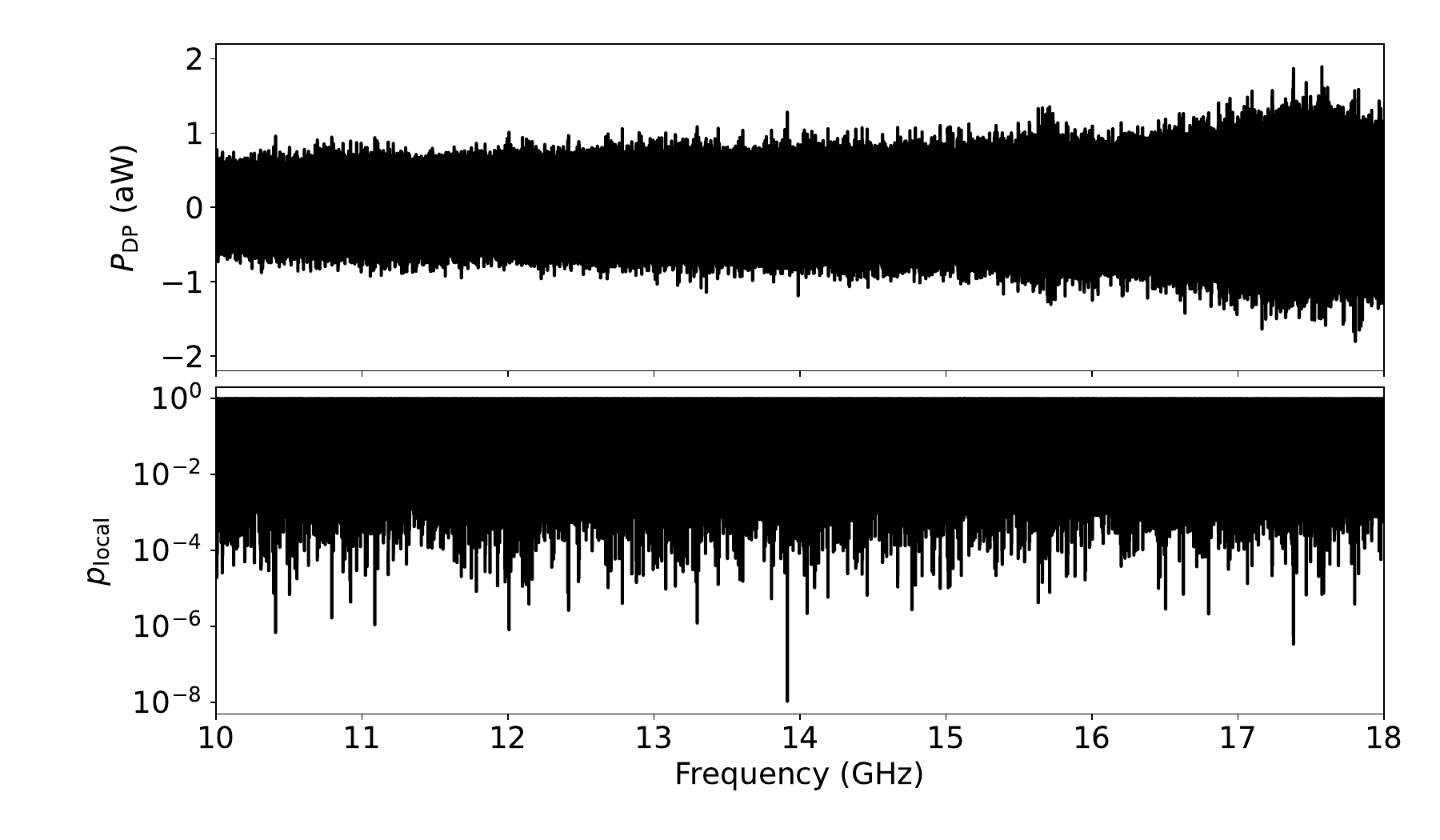}
\caption{\label{fig:result}
Extracted signal powers (top) and their local $p$ values (bottom) at each frequency in the initial data taken from March 6 to 17, 2023.
}
\end{figure}

\begin{figure}[htb]
\includegraphics[width=8.6cm]{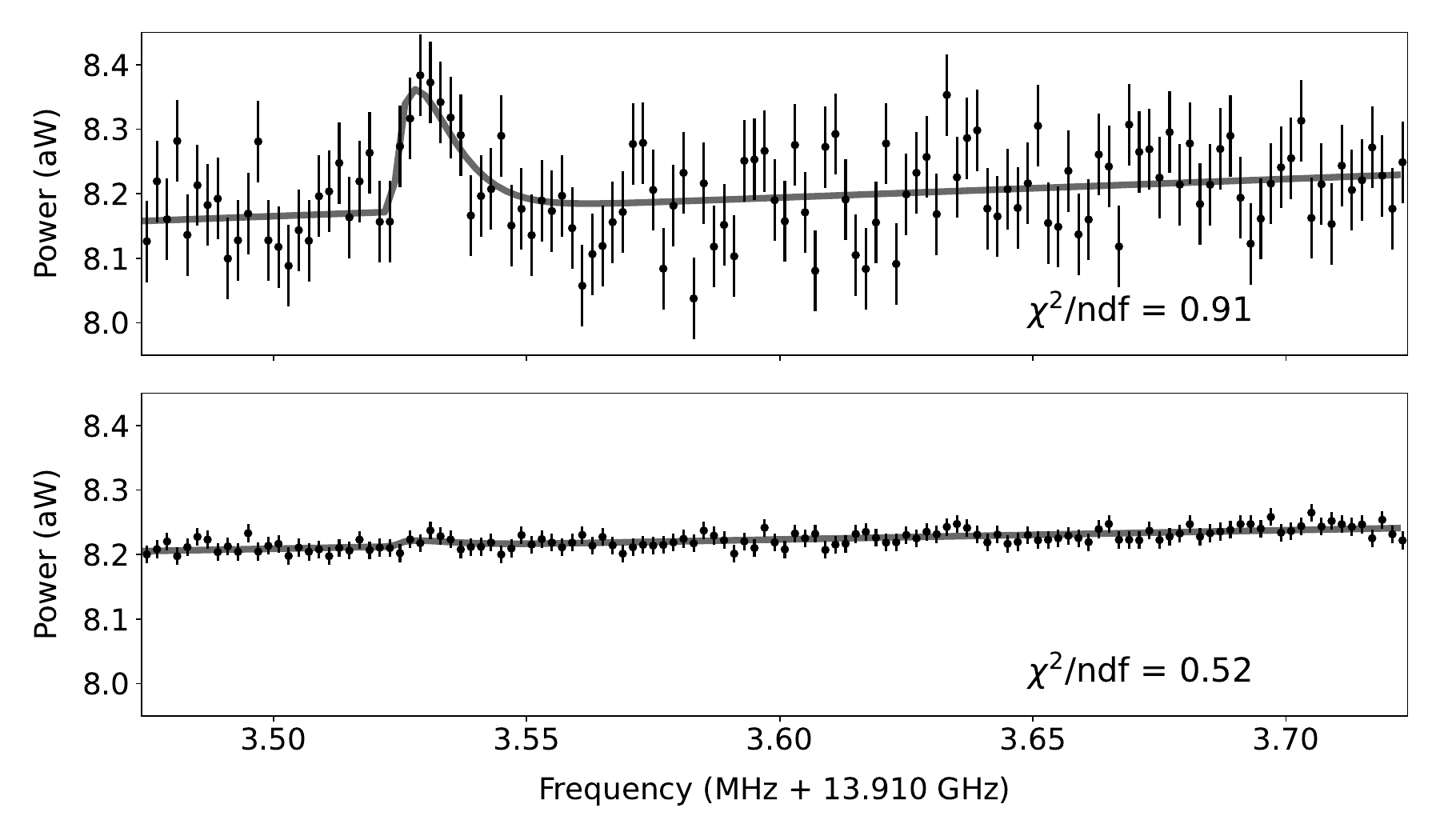}
\caption{\label{fig:excess}
Measured powers in the frequency region of the minimum $p$ value in the initial data.
Fit curves are overlaid.
The top panel shows the results of the initial data.
The bottom panel shows the results of the additional data whose statistics are 50 times higher than the initial data.
No significant signal was found in the data with increased statistics.
}
\end{figure}

\Fig{result} shows the fit results for the data. 
Over the entire frequency range, $\PDP$ was less than a few $\aW$.
We calculated the statistical significance at each frequency in the same manner as in our previous study~\cite{DOSUE-K}.
The minimum local $p$ value was $1.1\times 10^{-6}\,\%$ at 13.913524 GHz.
The spectrum at this frequency is shown in the top panel of \fig{excess}.
Furthermore, we calculated the global $p$ value 
(i.e., the probability of exceeding the minimum local $p$ value at any frequency) 
to be $2.2\,\%$, which is not a significant excess.
The calculation methods for the local and global $p$ values are described in \App{stats}.

\begin{figure}[htb]
\includegraphics[width=8.6cm]{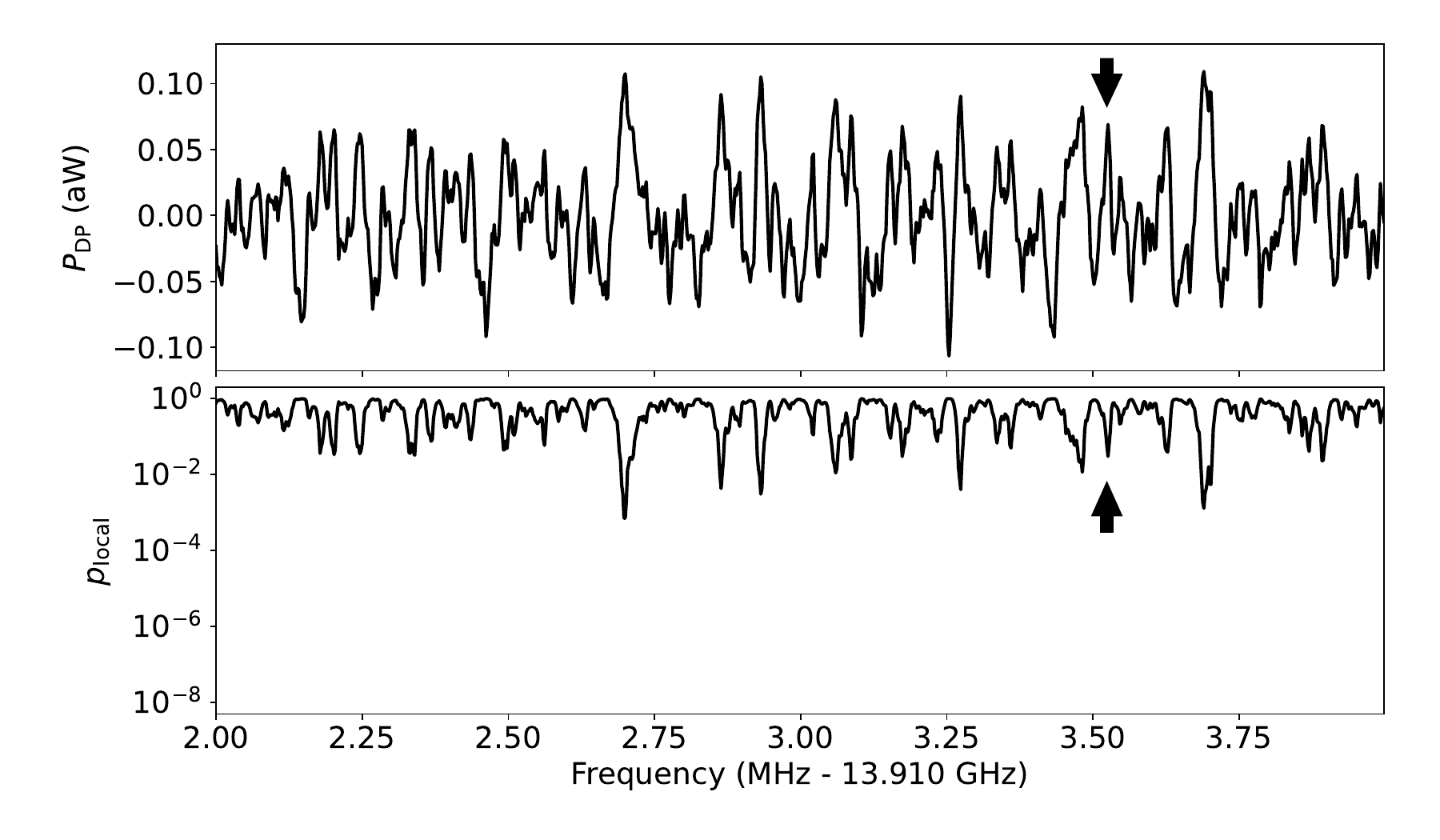}
\caption{\label{fig:result_add}
Extracted signal powers (top) and their local $p$ values (bottom) as a function of frequency for the additional data.
The arrows indicate the frequency of the minimum $p$ value in the initial data.
}
\end{figure}

To further verify the results at 13.913524 GHz,
we collected additional data focusing on the frequency range from 13.912 to 13.914 GHz on April 6, 2023.
We collected 600 data chunks, having 50 times more statistics than the initial data.
The spectrum and fit results at 13.913524 GHz are shown in the bottom panel of \fig{excess}.
The local $p$ value is 5.7\%.
The $\PDP$ and local $p$ values in this frequency range are shown in \fig{result_add}.
We confirmed that there was no significant excess.
Even after combining initial and additional data, 
the local $p$ value was 1.2\%.
Therefore, we concluded that no significant signal exists.

\section{Systematics Uncertainties}\label{sec:syst}
\begin{table}[htb]
\caption{\label{tab:syst}
Systematic uncertainties associated with the coupling constant $\chi$.
The uncertainty of the antenna pointing direction depends on the frequency.
}
\begin{ruledtabular}
\begin{tabular}{lr}
    Source                  & ($\%$) \\
    \colrule
    \textrm{Antenna-pointing direction}              & 0.3--7.8 \\
    \textrm{Effective aperture area ($\Aeff$)}       & 3.1 \\
    \textrm{Earth speed variation ($\vE$)}           & 2.3 \\
    \textrm{Frequency binning}                       & 1.3 \\
    \textrm{Gain}                                    & 1.1 \\
    \textrm{Alignment of instruments}                & $<$ 0.1 \\
    \textrm{Direction of conversion photons}         & $<$ 0.1 \\
    \textrm{Dark matter density ($\rho$)}            & 3.9 \\
    \colrule
    \textrm{Total}                                   & 5.8--9.7 \\  
\end{tabular}
\end{ruledtabular}
\end{table}
The systematic uncertainties associated with the coupling constant $\chi$ are summarized in \tab{syst}. 
Uncertainty from antenna-pointing direction 
was estimated from the beam center shift from $0\,\Deg$
as mentioned in \sect{Aeff}.
The beam center shifts decrease the measured power of the conversion photons.
These effects were estimated assuming a Gaussian beam pattern.
Larger errors were assigned above $17\,\GHz$, as shown in the \fig{beamcenter}.
The uncertainty in $\Aeff$ was estimated from the difference between the calibration and simulation results, as shown in \fig{Aeff}.
The variation in Earth speed $\vE$ used in \eq{fitF} during the search measurements creates a small bias in $\PDP$.
This was estimated by changing $\vE$ used in the fit for the DPDM signal simulation. 
Frequency binning causes another bias in the fit.
This was estimated from the variation in the $\PDP$ when the signal frequency was varied by $\pm\Dnu/2=\pm1\,\kHz$ in the simulation.
Some of the uncertainties in the gain were conservatively considered by 
the maximum difference between the two calibrations, as mentioned in \sect{measurement} (1.5\% on $\PDP$ corresponds to 0.8\% on $\chi$).
Furthermore, gain uncertainties from the source temperature and emissivity were considered (0.7\% on $\chi$).
The square root of the sum of their squares was assigned to the total systematic error of the gain.
For instrumental alignment, the tilts of the plate and antenna were at most $0.1\Deg$ and $0.05\Deg$, respectively.
This effect was estimated using the same method as that used for the uncertainty from the antenna-pointing direction.
Similarly, the uncertainty related to the direction of the conversion photons was estimated to be less than $0.1\%$.
For the dark matter density, we used the uncertainty described in \cite{DMdensity}. 
The total systematic error was $\text{5.7--9.7}\%$.

\section{Constraints on the Coupling Constant}\label{sec:limit}
\begin{figure}[htb]
\includegraphics[width=8.6cm]{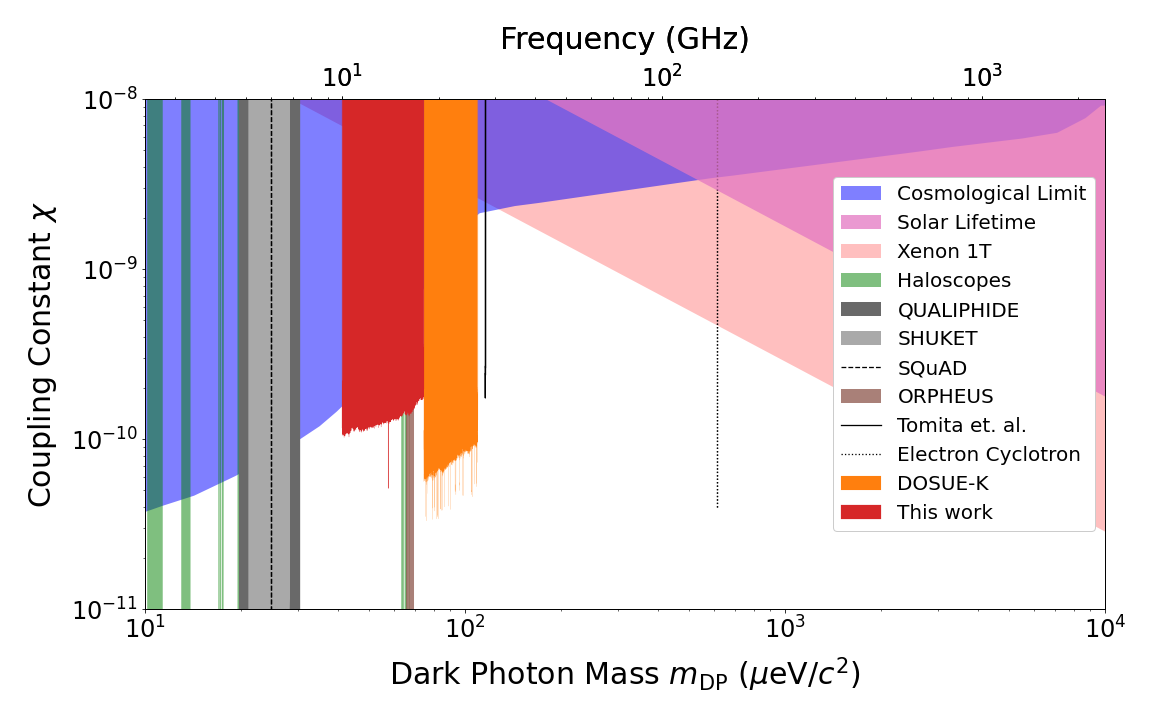}
\caption{\label{fig:limit}
Upper limits to the coupling constant $\chi$ at 95\% confidence level.
~\cite{DOSUE-K, QUALIPHIDE, SHUKET, SQuAD, ORPHEUS, Tomita, ElectronCyclotron, Wispy, SolarLifetime, Xenon1TLightDM, ORGAN, AxionLimit} 
}
\end{figure}

The upper bounds of $\PDP$ at a 95\% confidence level for each frequency were calculated
in the same manner as in \cite{DOSUE-K}.
The details are provided in \App{stats}.
The upper limits on $\PDP$ combining the initial and additional data, were converted into the upper limits on $\chi$ using \eq{chi}.
Systematic uncertainties were considered during this process.
As shown in \fig{limit}, we obtained the limits for $\chi < (\text{0.5--3.9}) \times 10^{-10}$ at a 95\% confidence level in the mass range $\text{41--74}\,\ueV$.
This is tighter than that provided by the cosmological observations~\cite{Wispy}. 
In a different mass range, our previous study which used a cryogenic system already achieved tighter constraints than this study~\cite{DOSUE-K}.

\section{Conclusions}
We searched for DPDM using a room-temperature millimeter-wave receiver and radioshielding box. The explored frequency range $\text{10--18}\,\GHz$ corresponds to the mass range $\text{41--74}\,\ueV$. We found no significant signal and set the upper limits of $\chi < (0.5\text{--}3.9) \times 10^{-10}$ at the 95\% confidence level. The achieved constraint was tighter than that obtained from cosmological observations.

\section*{Acknowledgements}
Part of this work was supported by Microwave Energy
Transmission Laboratory (METLAB), Research Institute for
Sustainable Humanosphere, Kyoto University.
The authors thank Naoki Shinohara and Hideki Ujihara.
This research was supported by JSPS KAKENHI under grant numbers 
20K14486,
20K20427,
21H01093,
21H05460,
and
23H00111.
This research was also supported by grants from the Murata and Sumitomo Foundations.
SA and TS acknowledge the Hakubi Project and SPIRITS Program of Kyoto University, respectively.

\appendix
\section{Statistical Calculations}\label{app:stats}
\subsection{Local $p$ value}\label{app:localp}
\begin{figure}[htb]
\includegraphics[width=8.6cm]{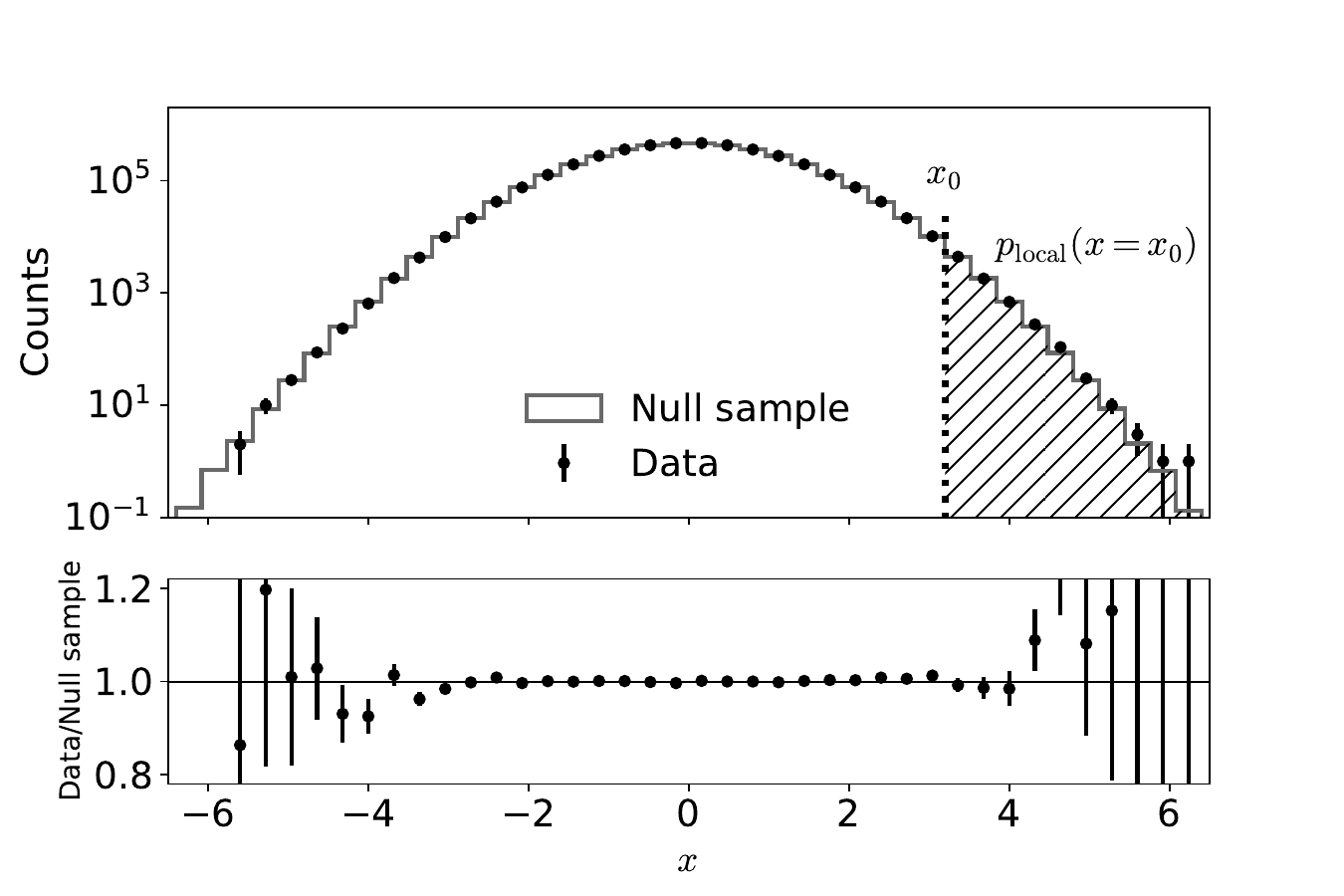}
\caption{\label{fig:null_hist}
Distributions of $x\equiv P_\mathrm{DP}/\sigma$ for data and null samples($\mathcal{P}(x)$).
Here, the plot for the null samples was normalized to be the same area as the data.
The local $p$ value for the fit result with $x=x_0$ was obtained from the fraction of the shaded area.
}
\end{figure}

To understand the statistical significance of the fit results for $\PDP$, we employed the null-sample method.
This method was used in our previous study~\cite{DOSUE-K}.
The null samples are the spectra generated by subtracting six data chunks from the other six data chunks 
in the measured 12 data chunks, which model the noise-only spectra.
There were 462 combinations (i.e., $\binom{12}{6}\times 1/2$) generating null samples at each frequency range.
For each null sample at each frequency, we obtained $\PDP$, error $\sigma$, and their ratio $x\equiv \PDP/\sigma$.
\Fig{null_hist} shows the distribution of the $x$ obtained from the fit to the data and null samples.
The null-sample distribution $\mathcal{P}(x)$ is consistent with the data.
We calculated a local $p$ value under the zero-signal hypothesis for the fitted result with $x=x_0$ as follows:
\begin{equation}\label{eq:plocal}
 p = \frac{ \int_{x_0}^{\infty}{\mathcal{P}(x)}d x }{ \int_{-\infty}^{\infty}{\mathcal{P}(x)}d x }~.
\end{equation}
The fraction of the shaded area in \fig{null_hist} corresponds to this value.

\subsection{Global $p$ value}\label{app:globalp}
\begin{figure}[htb]
\includegraphics[width=8.6cm]{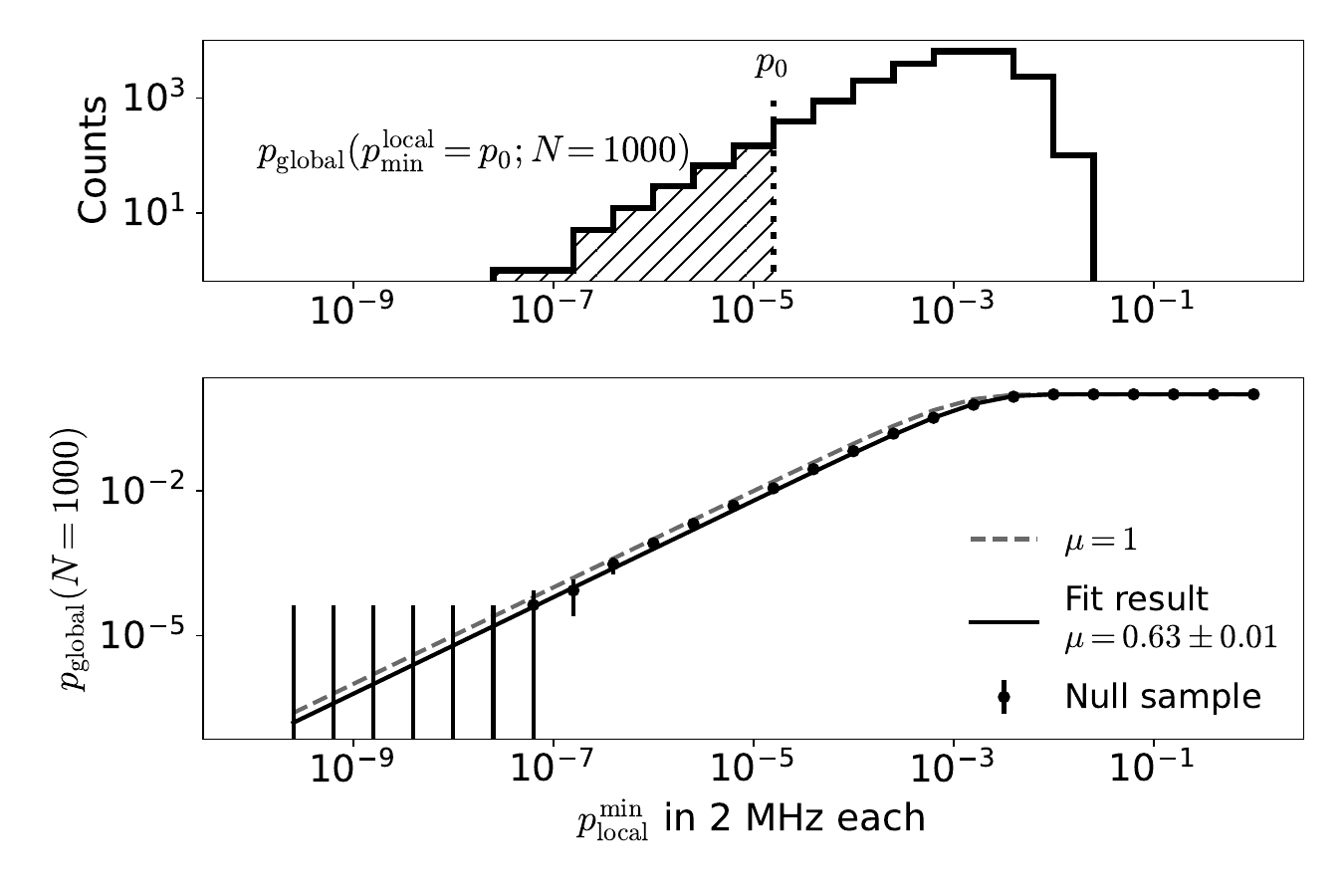}
\caption{\label{fig:pglobal}
$\plocalmin$ distribution (top) when a number of fittings $N$ is $1000$ and a fitting of global $p$ value (bottom) with a free parameter of $\mu$ in the frequency range $\text{10.0--10.1}\,\GHz$.
The fraction of the shaded area in the top panel is an estimated global $p$ value of the probability that the minimum local $p$ value is $p_0$ when $N=1000$.
}
\end{figure}

There are many fitting results in the frequency range 10--18 GHz ($4\times 10^6$).
The global $p$ value, which considers the look-elsewhere effect, is important for determining whether a signal signature exists.
Using the minimum $p$ local $\plocalmin$, the global $p$ value for $N$ fittings was calculated using the following formula~\cite{Tomita, AxionGlobalP}:
\begin{equation}\label{eq:pglobal}
    \pglobal(\plocalmin; N) = 1 - (1- \plocalmin)^{\sum_{i=0}^{N} \mu_i},
\end{equation}
where $\mu_i$ ($0< \mu_i \leq1$) is a scale factor for $i$-th fitting that accounts for the independence between the fittings in the neighbor bins.
Because the peak width of the DPDM signal depends on the frequency, $\mu_i$ depends on the frequency.
To estimate $\mu$ for this experiment, we separated the null samples into 2 MHz frequency intervals, each of which contained 1000 fit results, 
and checked the distribution of the minimum local $p$ values ($\plocalmin$) in each of the 2 MHz intervals.
The $\plocalmin$ distribution for the frequency range of $\text{10.0--10.1}\,\GHz$ is shown in the top panel of \fig{pglobal}.
A global $p$ value with $N=1000$ is calculated from the $\plocalmin$ distribution $\mathcal{Q}(\plocalmin)$ as follows:
\begin{equation}\label{eq:pglobal2}
 \pglobal(\plocalmin _{0}; N) = \frac{ \int_{0}^{ \plocalmin _{0} }{\mathcal{Q}(\plocalmin;N)}d \plocalmin }{ \int_{0}^{\infty}{\mathcal{Q}(\plocalmin;N)}d \plocalmin }~.
\end{equation}
By using the plot of the calculated $\pglobal$ at each $\plocalmin$,
we estimated $\mu$ within this frequency range, as shown in the bottom panel of \fig{pglobal}.

\begin{figure}[htb]
\includegraphics[width=8.6cm]{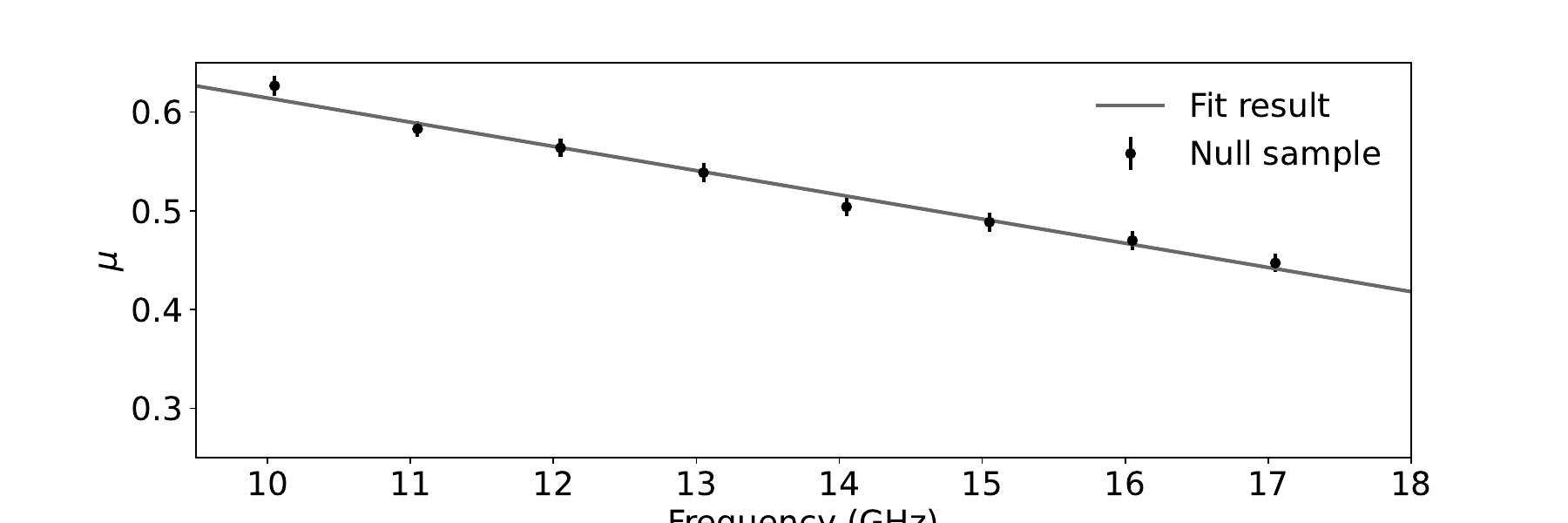}
\caption{\label{fig:mu}
Estimated $\mu$ for each frequency and the linear fit.
}
\end{figure}

We extracted $\mu$ for each frequency range of 10.0--10.1, 11.0--11.1 GHz,..., and 17.0--17.1 GHz as shown in \fig{mu}.
By fitting with a linear function, $\mu$ at $\nu_i$ can be obtained as 
\begin{equation}\label{eq:mu}
    \mu(\nu_i) = (-0.025\pm 0.001) \nu_i \text{[GHz]} + (0.86\pm0.02),
\end{equation}
where $\nu_i$ is the signal peak frequency at the $i$-th fit.
We calculated the global $p$ value using \eq{pglobal} and \eq{mu}.

\subsection{Upper Limits on $\PDP$}
Using \eq{plocal}, we obtained $p=0.05$ at $x_0=1.79$, 
which is slightly larger than that of a normal Gaussian distribution (1.65~\cite{pdg}).
This is because the distribution tail of $\mathcal{P}(x)$ is slightly wider than the normal Gaussian.
The upper bounds on $\PDP$ at the 95\% confidence level for each frequency were calculated as
\begin{equation}\label{eq:PDPlimit}
    \text{max}(0, \PDP) + 1.79 \sigma~.
\end{equation}


%

\end{document}